\documentclass[twocolumn]{aastex631}

\usepackage{natbib}
\usepackage{booktabs}

\begin{document}
\title{Testing the Association of Supermassive Black Hole Infrared Flares and High-energy Neutrinos}

\author{Megan Wang}
\affiliation{MIT Kavli Institute for Astrophysics and Space Research, Massachusetts Institute of Technology, Cambridge, MA 02139, USA}

\author[0009-0001-9034-6261]{Christos Panagiotou}
\affiliation{MIT Kavli Institute for Astrophysics and Space Research, Massachusetts Institute of Technology, Cambridge, MA 02139, USA}

\author[0000-0002-8989-0542]{Kishalay De}
\affiliation{Department of Astronomy and Columbia Astrophysics Laboratory, Columbia University, 550 W 120th St. MC 5246, New York, NY 10027, USA}
\affiliation{Center for Computational Astrophysics, Flatiron Institute, 162 5th Ave., New York, NY 10010, USA}

\author[0000-0003-0172-0854]{Erin Kara}
\affiliation{MIT Kavli Institute for Astrophysics and Space Research, Massachusetts Institute of Technology, Cambridge, MA 02139, USA}

\author[0000-0003-4127-0739]{Megan Masterson}
\affiliation{MIT Kavli Institute for Astrophysics and Space Research, Massachusetts Institute of Technology, Cambridge, MA 02139, USA}

\author[0000-0002-0525-3758]{Foteini Oikonomou}
\affiliation{Department of Physics, Norwegian University of Science and Technology, Høgskoleringen 5, Trondheim 7491, Norway}

\correspondingauthor{Christos Panagiotou}
\email{cpanag@mit.edu}

\begin{abstract}
The physical origin of the observed cosmic neutrinos remains an open question and the subject of active research. While matter accretion onto supermassive black holes is long thought to accelerate particles to high energies, it has recently been suggested that tidal disruption events, and accretion flares in general, with prominent IR echoes can account for a fraction of the diffuse high-energy neutrino signal. Motivated by this result, we compile a sample of nearby accretion flares detected in the NEOWISE survey featuring strong IR echoes, and we cross-match it with the latest catalog of neutrino alerts, IceCat-1. We recover only a
single spatial coincidence between the two catalogs, consistent with a chance coincidence. We find no temporal and spatial coincidences between the two samples, which, given the properties of our sample, 
appears to challenge previous conclusions. We discuss the physical implications of our results and potential future explorations.

\end{abstract}

\section{Introduction}

Observations of astrophysical neutrinos offer a novel and independent way to probe high-energy astrophysical phenomena. 
The IceCube Neutrino Observatory began operations in 2011, detecting neutrinos and providing a new view of the multimessenger sky. Data collected revealed a diffuse high-energy neutrino flux, consistent with a neutrino population that has origins beyond the solar system \citep{IceCubeCollab2013, IceCubeCollab2024}. The sources that contribute to this astrophysical neutrino background, however, remain largely unknown. 

Accretion of matter onto supermassive black holes, which powers active galactic nuclei (AGN), comprises one of the most efficient ways to release energy and has been proposed as a mechanism to accelerate particles to high energies \citep{Farrar2009, Murase2017}. In fact, NGC 1068, a nearby radio-quiet Seyfert galaxy, is so far the only extragalactic source found to emit high-energy neutrinos with high significance, above 4$\sigma$ \citep{IceCubeCollab2022}. A few recent studies have also reported evidence for neutrino emission from three additional nearby Seyfert galaxies, namely NGC 4151, NGC 3079, and CGCG 420-015, further strengthening the link between accreting supermassive black holes and detected neutrino events \citep{Neronov2024, 2025ApJ...981..131A, 2025ApJ...988..141A}. Moreover, the blazar TXS 0506+056 was identified as a potential neutrino source after a neutrino event was shown to coincide temporally and spatially with a gamma-ray flare from this object \citep{IceCubeCollab2018a}. While the population of known blazars is not expected to dominate the TeV-PeV neutrino sky \citep{Aartsen2017}, this result motivated additional searches for a connection between neutrino signals and transient events.

If steady matter accretion onto supermassive black holes, as is the case in typical AGN, is a source of astrophysical neutrinos, it is tempting to explore whether episodic accretion events, such as tidal disruption events \citep[TDEs,][]{Rees1988, Gezari2021} or flaring AGN, may also contribute to the observed neutrino sky. 
TDEs occur when an unlucky star wanders too close to a black hole and gets shredded apart by tidal forces. A fraction of the stellar debris is accreted onto the black hole, leading to emission across the electromagnetic spectrum.

The first strong evidence for a connection between TDEs and neutrino events was presented by \cite{Stein2021}, who deduced the likely association of a spectroscopically classified TDE, AT2019dsg, with the IceCube neutrino event IC191001A, thanks to real-time IceCube alert follow-up by the Zwicky Transient Facility (ZTF). Later, \cite{Reusch2022} reported a coincidence between a neutrino event and a prominent AGN flare, AT2019fdr, which was classified as a TDE candidate. Building on these two results, \cite{vanVelzen2024} conducted a systematic search of ZTF nuclear flares and found a third coincidence between IceCube alerts and a TDE candidate, namely AT2019aalc. They estimated a significant association between high-energy neutrinos and accretion flares and concluded that around 20\% of the IceCube neutrino events may originate from accretion flares. However, it should be noted that a later detailed search utilizing the full neutrino data sample did not find a significant excess of neutrino signal in the direction of optical accretion flares \citep{Necker2023}. More recently, \cite{Jiang2023} presented two likely coincidences between neutrino events and TDE candidates, identified in a sample of mid-infrared outbursts in nearby galaxies, and \cite{2025ApJ...991...20J} reported a coincidence between a repeating TDE candidate and two neutrino events.

Interestingly, a common characteristic of all the TDE candidates linked with neutrino events, is a strong mid-IR echo. This mid-IR emission is a result of the surrounding dust reprocessing the original UV/optical flare \citep[e.g.][]{Jiang2016, vanVelzen2016, Panagiotou2023}. The reported coincidences also exhibit a neutrino arrival time delayed relative to the optical flare peak of the TDE candidates, but well aligned with the mid-IR peak, hinting at a potential connection between neutrino emission and dust echoes. Time-dependent models of neutrino production in TDEs have tried to theoretically explain this time delay due to possible delayed onset of particle acceleration \citep{Winter2023}.

Motivated by these results, we compile a catalog of bright nearby mid-IR accretion flares utilizing the  NEOWISE survey. In contrast to previous studies, our sample consists of accretion flares identified in the mid-IR wavelength range, with little to no optical counterpart, suggesting high levels of dust obscuration or intrinsic optical weakness \citep{Jiang2023}. These sources comprise an ideal data set to examine whether accretion flares with prominent dust echo are strong neutrino emitters, whereas they may have been missed in previous optical and X-ray surveys \citep{Masterson2024}. We use the latest neutrino catalog published by the IceCube collaboration to perform this exploration. 
The construction of our sample of mid-IR accretion flares is presented in Section \ref{sec:data}. The main results of our analysis are given in Section \ref{sec:results} and are discussed in Section \ref{sec:discussion}.

\section{Data} \label{sec:data}

\subsection{WISE Accretion Flare Selection} \label{subsec:flares}

The Wide-field Infrared Survey Explorer (WISE) satellite was launched in 2009 and was active until 2011. It was reactivated in 2013 as NEOWISE, surveying the entire sky every six months in two infrared wavebands, W1 (3.4 $\mu$m) and W2 (4.6 $\mu$m), until being deactivated in 2024.
In this work, we search the NEOWISE archive to construct a catalog of nearby accretion flares. Initially, we employ a customized pipeline to perform difference photometry on coadded data from the unWISE project to detect mid-infrared transients \citep{Lang2014,Meisner2017,Meisner2017b}, using for reference images the full-depth unWISE coadded images from 2010 to 2011. Further details of the process followed to identify transient events will be presented in a dedicated future publication \cite[De et al. in prep., see also][]{De2023Natur}. We, then, identify nuclear transients by spatially cross-matching our sample of NEOWISE transients to the Census of the Local Universe (CLU-compiled) catalog, a highly complete collection of known galaxies with spectroscopic redshift within 200 Mpc \citep{Cook2019}. More specifically, we require the transient event to occur within 2'' from the center of a galaxy. It should be mentioned that the host galaxies of the TDE candidates AT2019dsg, AT2019fdr and AT2019aalc are not in the CLU catalog and these events will thus not be included in our sample.

Following \cite{vanVelzen2024}, we seek to obtain a sample of high-amplitude, fast rising, and slowly declining accretion flares. However, the population of mid-infrared transients produced by the above cross-match contains our desired accretion flares as well as contaminants such as supernovae (SNe) and standard AGN featuring normal stochastic variability. 

To minimize contamination from SNe, we consider the duration of the transient event, defined as the temporal interval over which the event is detected over the host galaxy's emission in difference imaging.  We remove events with a duration shorter than two years, since typical SNe are not expected to be bright in the mid-IR for longer than that \citep[e.g.][]{Wright1980, Johansson2017}, while accretion flares typically last considerably longer \citep{Masterson2024, Necker2025}. We confirmed a posteriori that most optical flares of \cite{vanVelzen2024}, up to 80\%, including the three with neutrino association, satisfy this criterion as well.

Then, we consider the complete NEOWISE W1 light curves of our sources in order to eliminate AGN displaying typical stochastic variability. To compute the W1 light curves, we performed aperture photometry on the estimated position of each transient and applied the WISE photometric zero points to obtain the flux in physical units  \citep{Wright2010}\footnote{\url{https://wise2.ipac.caltech.edu/docs/release/allsky/expsup/sec4_4h.html}}. Using these light curves,
we first remove sources with low variability by calculating the fractional rms variability amplitude   \citep[$F_{var}$,][]{Vaughan2003} of our light curve and requiring $F_{var} \geq 0.1$. We further remove transients that show only a minimal increase in brightness to keep solely sources featuring a prominent flare in our sample. To enforce this, we estimate the average peak magnitude of a flare as the mean magnitude within one year of the brightest observation.
We also define the baseline magnitude as the average pre-flare magnitude of the source. Then, we require a magnitude decrease of $\Delta m < -0.2$ between the baseline and peak magnitude of the flare, which is typically a lower threshold for IR selected TDEs \citep{Masterson2024} and larger than regular AGN variability on the considered time scales \citep[e.g.][]{2023ApJ...958..135S}.

\begin{figure}
    \centering
    \includegraphics[width=1\linewidth]{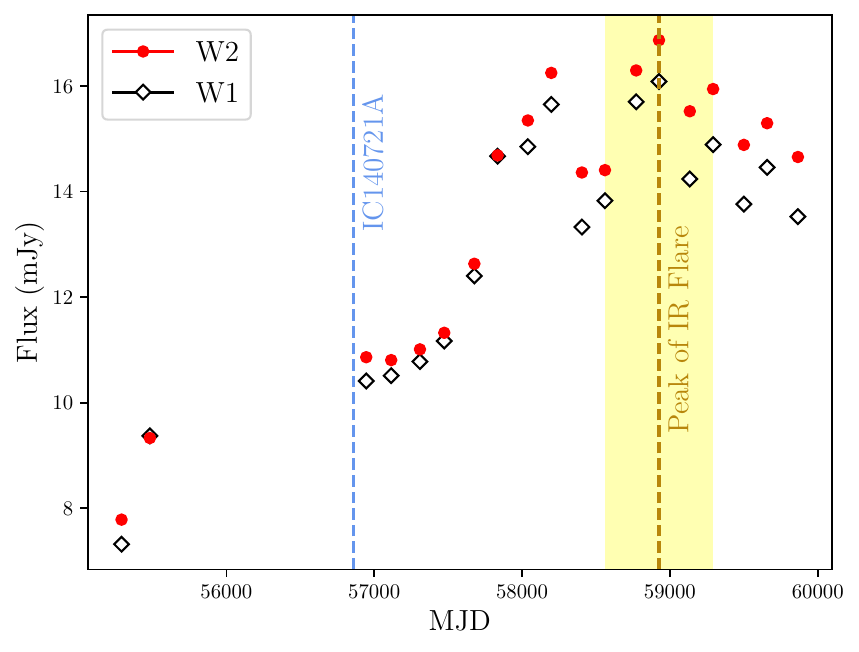}
    \caption{The NEOWISE W1 (3.4$\mu m$) and W2  (4.6$\mu m$) aperture photometry light curve of the accretion flare WTP14aczncp, which was found to be spatially coincident with the neutrino event IC140721A. The arrival window for the employed temporal coincidence criterion is highlighted with yellow, and the neutrino arrival time is marked with a blue dashed line. 
    Open black diamonds and filled red circles represent the W1 and W2 waveband flux respectively.}
    \label{fig:light_curves}
\end{figure}

Altogether, these criteria yield a final sample of 99 accretion flare candidates, consisting of both TDEs and high-amplitude flares from AGN. An example light curve of one such transient is shown in Fig. \ref{fig:light_curves}, while the details of all our sources are listed in Table \ref{table:flares}.
It should be noted that a number of the transients in our sample have previously been presented as TDE candidates in \cite{Masterson2024}. Our sample construction successfully recovers all high-confidence TDEs of that work.

\subsection{IceCube Neutrino Alerts}
The IceCube Neutrino Observatory \citep{Abbasi2009}, located at the South Pole, detects Cherenkov radiation produced by secondary charged particles that result from neutrino interactions with the ice. Our parent sample of neutrino alerts is the most recent catalog of astrophysical neutrino events produced by the IceCube collaboration, IceCat-1 \citep{Abbasi2023}. This catalog consists of track-like events which are mainly produced by muons traveling through the instrumented volume of IceCube. The direction and energy of track-like events can then be reconstructed and used to estimate the likelihood of an event being astrophysical in origin. This likelihood is a function of the reconstructed energy, reconstructed declination, and assumed astrophysical neutrino spectrum. The catalog contains events from May 13, 2011 to December 31, 2020. For this analysis, we only consider events following January 1, 2014 because the reactivation of the WISE satellite as the NEOWISE mission does not occur until September 2013. Moreover, we only consider 
neutrino events with signalness above 0.5, suggesting a larger than 50\% probability of being astrophysical \citep[see][for details]{Abbasi2023}. This signalness based criterion is similar to that of \cite{Jiang2023}, but stricter than the one employed in the exploration of \cite{vanVelzen2024}. We also exclude neutrino events with a 90\% angular uncertainty greater than 50$^{\circ}$ in any direction, resulting in the exclusion of one neutrino event, IC140410A. These criteria yield a final sample of 68 IceCube neutrino events. Together with our selection of 99 accretion flares, we use larger samples than previous works \citep[e.g.][]{Reusch2022, vanVelzen2024}, allowing a more detailed investigation of possible associations.

\section{Results} \label{sec:results}

\begin{figure*}
    \centering
    \includegraphics[width=1\linewidth]{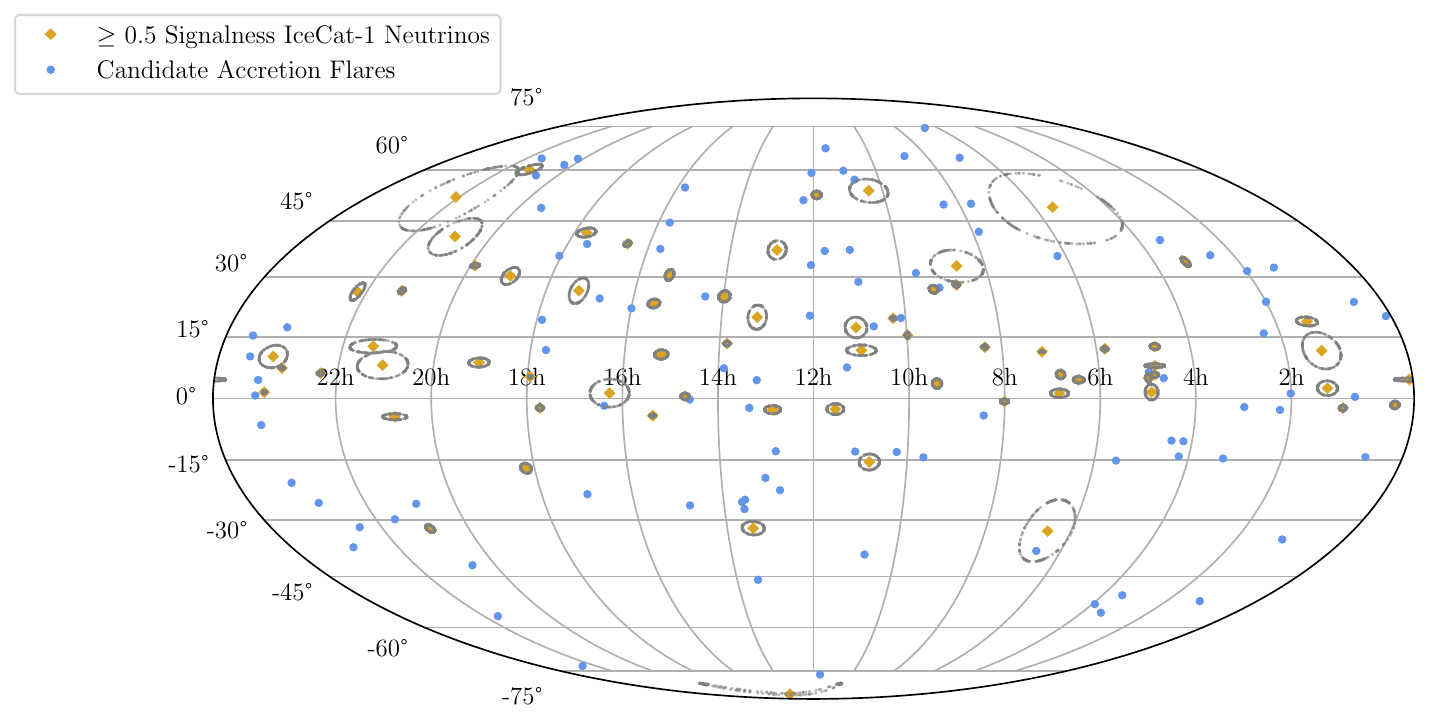}
    \caption{
    An all-sky map in equatorial coordinates of the considered IceCat-1 neutrino events, with signalness above 0.5 and 90\% angular uncertainty contours below 50$^\circ$as reported by \cite{Abbasi2023}, and our final sample of 99 mid-IR accretion flares.}
    \label{fig:skymap}
\end{figure*}

We cross-match our sample of accretion flare candidates, listed in Table \ref{table:flares}, with the neutrino sample from IceCat-1, in order to determine whether any of our mid-IR-selected transients are associated with neutrino events. A skymap of the accretion flare sample and all the considered IceCat-1 neutrino events is plotted in Fig. \ref{fig:skymap}.
We note that the two samples have different observational selection effects. An apparent excess of neutrino events in the northern hemisphere is due to the declination-dependent sensitivity of the IceCube observatory \citep[e.g.][]{2009PhRvL.103v1102A}. Instead, the NEOWISE survey provides a more homogeneous sensitivity across the whole sky, while the chosen galaxy catalog, CLU, also includes more galaxies in the northern hemisphere. These observing limitations become important when one wishes to estimate the exact fraction of astrophysical neutrinos a class of celestial sources can account for. However, these effects are not expected to affect significantly our analysis and main conclusions, as presented below. 

For a flare to be considered spatially coincident with a neutrino event, we require that the flare falls within the neutrino alert's 90\% uncertainty contours. In addition, for the neutrino event and accretion flare to be physically linked, the two events must also coincide temporally. We follow a conservative approach and assume a temporal coincidence when the neutrino arrives within one year of the transient's IR flux peak. This is consistent with past studies which found that for accretion flares coincident with neutrinos, the neutrinos have arrived in a narrow window around the IR peak of the flare \citep{Jiang2023,vanVelzen2024}. 

Our analysis leads to a singular spatial coincidence between WTP14aczncp and IC140721A. The light curve of the galactic nucleus hosting WTP14aczncp is shown in Fig. \ref{fig:light_curves} along with the neutrino event time. WTP14aczncp features two flares, both of which though are temporally far from the arrival time of IC140721A. While we lack WISE data in the years before the neutrino recorded time, 
the neutrino event seems to coincides with a low flux state for the IR transient, before the onset of the flare, which disfavors a physical association between the two events and points at a chance association. We deduce that our analysis yields zero complete coincidences between our sample of mid-IR accretion flares and high signalness neutrinos.

To further explore the lack of coincidence between accretion flares and neutrino events, and to conclude on its significance, we calculate the infrared W1 flux of our sources, which has been assumed to scale with the neutrino flux of accretion flares. 
The peak flux and baseline flux of each transient are estimated following the procedure described in \ref{subsec:flares}. In order to allow for a direct comparison with the work of \cite{vanVelzen2024}, we follow their approach in estimating the IR flux of a transient, $\Delta F_{IR}$, as the peak minus the baseline flux. The computed values are given in Table \ref{table:flares}. We further calculate the IR flux of all accretion flare candidates in \cite{vanVelzen2024} using our custom pipeline to estimate their NEOWISE W1 light curve. In this way, we can compare our sample with other accretion flares in the nearby Universe, including the three reported coincidences. Similarly, we measure the IR flux of the two TDE candidates considered in \citep{Jiang2023}.

\begin{figure}
    \centering
    \includegraphics[width=1\linewidth]{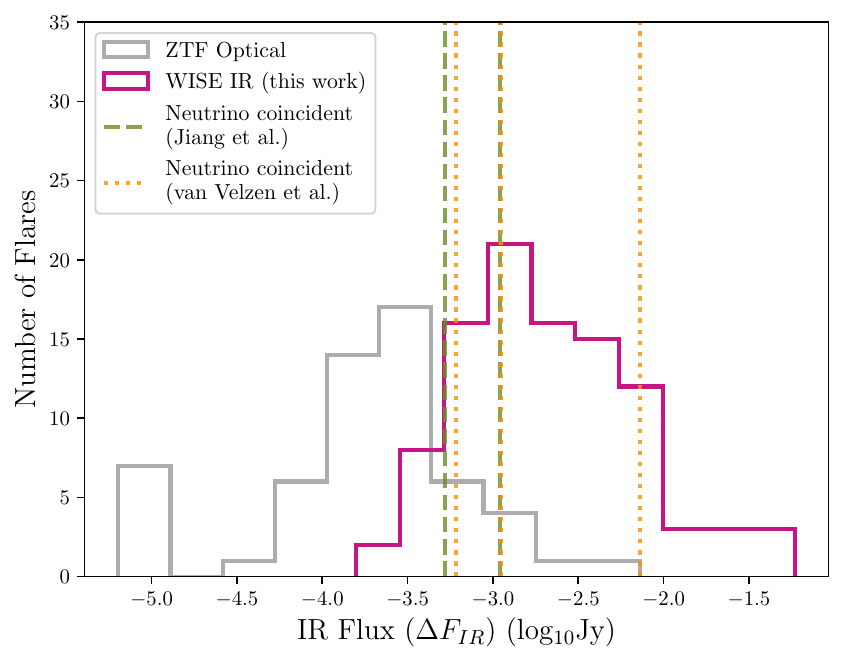}
    \caption{Distribution of the IR flux of the mid-IR accretion flares identified in our analysis (magenta) and of the optical accretion flares (gray curve) considered by \cite{vanVelzen2024}.The vertical orange dotted lines and green dashed lines denote the flux of the TDE candidates reported to coincide with a neutrino even by \cite{vanVelzen2024} and \cite{Jiang2023} respectively.}
    \label{fig:echo_calcs}
\end{figure}

Figure \ref{fig:echo_calcs} plots the histogram of the computed IR fluxes for all sources. It is evident that the mid-IR detected accretion flares, identified in our analysis, feature a systematically larger flux when compared to accretion flares identified in the optical. Their flux is also similar, if not larger, to the reported accretion flares with a neutrino event association.
Hence, the lack of any coincidence with neutrino events seems to challenge previous results that accretion flares with strong dust echo accounts for a fraction of the observed high-energy neutrino sky. In fact, \cite{vanVelzen2024} assumed that mid-IR emission, presumed to be a proxy for the bolometric flux, and neutrino flux of accretion flares are correlated in order to infer a statistically significant coincidence between the two populations. Under such an assumption, we expect our sample, given its size and the high mid-IR fluxes, to coincide with several IceCube events, which we do not observe. Therefore, we conclude that the general population of accretion flares does not feature a prominent neutrino emission.

To explore potential biases imposed by our selection criterion for the flare duration, we also investigate possible coincidences between high signalness neutrinos and shorter duration flares. Upon including all events with a duration greater than half of a year (i.e. events with more than 1 NEOWISE detection) in our analysis, and applying the same criteria outlined above, we again find no complete coincidences between these shorter duration flares and high signalness neutrinos, which further strengthens our results.

\section{Conclusions} 
\label{sec:discussion}

In this work, we compiled a sample of the brightest nearby mid-IR accretion flares by systematically searching for nuclear transients in NEOWISE survey data and enforcing additional criteria to select TDE candidates and high-amplitude AGN flares. Then, we utilized the latest IceCube neutrino catalog, IceCat-1, and we cross-matched our sample with neutrino events of high signalness to identify mid-IR accretion transients that are potentially sources of astrophysical neutrinos. Our analysis yields zero coincidences between accretion flares and astrophysical neutrinos, pointing at the lack of a prominent physical association between the two populations.

In retrospect, this is perhaps expected. As discussed in detail by \cite{vanVelzen2024}, the energy reservoir and the fluence of regular AGN is significantly larger than in transient accretion flares. As both phenomena are thought to be powered by the same process, that is the accretion of matter onto a supermassive black hole, neutrino emission of AGN should dominate the observed neutrino sky over any contribution from flaring events. Indeed, \cite{Fiorillo2025} have recently shown that AGN emission can account for the diffuse neutrino signal of IceCube, further confirming earlier explorations \citep[e.g.][]{Padovani2024}. In addition, preliminary results from a refined analysis of the IceCube events with improved angular resolution dispute the previously reported association between neutrino events and optical TDEs \citep{Zegarelli2025}.

Although our results disfavor a strong association between the general population of accretion flares and neutrino events, it should be mentioned that an association with a specific class of accretion flares cannot be ruled out. For example, if there are intrinsic differences between mid-IR and optical detected flares, perhaps only the latter could accelerate neutrinos efficiently. Nevertheless, our work highlights that accretion flares with strong dust echoes fail to account for a large fraction of the observed neutrinos, disputing a correlation between their mid-IR flux and neutrino emission. We motivate further work with larger samples to constrain the fraction of neutrino events, if any, associated with different groups of accretion flares. Future studies with the Rubin observatory \citep{2019ApJ...873..111I}, the Roman Space Telescope \citep{2019arXiv190205569A}, SPHEREx \citep{Dore2014arXiv}, and  NEOSurveor \citep{Mainzer2023} have the potential to conclusively address this.

\bibliography{ref}{}
\bibliographystyle{aasjournal}

\pagebreak
\centering
\setlength\LTleft{0pt} 
\setlength\LTright{0pt} 
\scriptsize
\noindent
\begin{longtable}{lrrrr}
\caption{WISE IR Transients. The flare names follow the naming scheme outlined in \cite{2023MNRAS.523.3555Z}. The fourth columns lists the peak W1 flux of the transient event. The redshift values were retrieved from NED\footnote{\url{https://ned.ipac.caltech.edu/}}.} \\
\toprule
\textbf{Flare Name} &\textbf{Right Ascension} &\textbf{Declination} &\textbf{W1 flux (mJy)} &\textbf{Redshift} \\
\midrule
\endfirsthead
\caption[]{WISE IR Transients (continued)} \\
\toprule
\textbf{Flare Name} &\textbf{Right Ascension} &\textbf{Declination} &\textbf{W1 flux (mJy)} &\textbf{Redshift} \\
\midrule
\endhead
\midrule
\endfoot
\bottomrule
\endlastfoot
WTP14aczncp & 102.573 & -38.087 & 6.9967 & 0.0300 \\
WTP18aamodk & 36.961 & 1.254 & 0.1578 & 0.1863 \\
WTP14acrehd & 180.857 & 33.146 & 0.3424 & 0.0349 \\
WTP14acrprg & 346.762 & 4.549 & 6.4714 & 0.0430 \\
WTP14abmntz & 57.723 & 39.789 & 4.1231 & 0.0187 \\
WTP20aancak & 79.111 & 6.472 & 16.6613 & 0.0339 \\
WTP17aamhvw & 74.789 & 4.975 & 1.9970 & 0.0156 \\
WTP14acqmec & 146.425 & -14.326 & 44.4391 & 0.0077 \\
WTP17aaiosn & 304.297 & 58.202 & 2.5429 & 0.0123 \\
WTP14acyrpi & 103.049 & 74.427 & 28.5947 & 0.0195 \\
WTP14aclcgv & 54.778 & -14.611 & 1.2554 & 0.0358 \\
WTP14adccll & 11.386 & -14.259 & 0.8398 & 0.0220 \\
WTP10aczqiu & 199.264 & -2.261 & 1.2763 & 0.0193 \\
WTP14acqreo & 161.390 & 17.668 & 0.8511 & 0.0414 \\
WTP20aajfjy & 200.861 & -45.970 & 1.5558 & 0.0366 \\
WTP14aczyhc & 128.902 & 49.595 & 1.3211 & 0.0424 \\
WTP17aamoxe & 317.243 & -56.475 & 4.5237 & 0.0431 \\
WTP14acljhb & 44.697 & 35.733 & 3.3690 & 0.0454 \\
WTP13aabtfk & 167.311 & -12.931 & 1.6626 & 0.0256 \\
WTP15acbuuv & 154.673 & -13.002 & 5.7886 & 0.0306 \\
WTP15abzydu & 266.572 & 35.572 & 5.1269 & 0.0225 \\
WTP14abnpgk & 251.478 & -23.452 & 7.2920 & 0.0203 \\
WTP14acxcsm & 259.161 & 38.724 & 2.4499 & 0.0728 \\
WTP14adaukh & 330.508 & -31.870 & 58.5813 & 0.0087 \\
WTP14acstok & 1.581 & 20.203 & 20.3012 & 0.0258 \\
WTP14adbrmc & 176.125 & 36.886 & 2.1696 & 0.0381 \\
WTP18aanzwl & 152.783 & 19.741 & 0.7327 & 0.0289 \\
WTP14ackyvh & 41.778 & 15.929 & 1.4641 & 0.0250 \\
WTP15abzcdt & 36.727 & 23.800 & 6.2325 & 0.0336 \\
WTP14adbonj & 184.029 & 50.825 & 1.8774 & 0.0310 \\
WTP16aaqrcr & 280.830 & -62.112 & 3.0773 & 0.0151 \\
WTP15abulxt & 68.453 & -14.090 & 0.6949 & 0.0310 \\
WTP14acszcf & 234.160 & 54.559 & 8.6476 & 0.0389 \\
WTP15abnrzj & 96.971 & 35.517 & 2.9096 & 0.0250 \\
WTP18aalxpd & 194.938 & -19.404 & 0.4928 & 0.0457 \\
WTP20aajikb & 71.658 & -10.226 & 3.4916 & 0.0151 \\
WTP14actuee & 202.239 & -27.225 & 1.5414 & 0.0448 \\
WTP14aczsef & 133.803 & 64.396 & 3.2934 & 0.0362 \\
WTP14ackhvn & 8.958 & 23.744 & 1.6797 & 0.0675 \\
WTP15abwllz & 302.891 & -41.959 & 0.9904 & 0.0486 \\
WTP14acqidh & 338.671 & -37.112 & 1.2299 & 0.0430 \\
WTP14actpwn & 201.803 & -24.862 & 1.7117 & 0.0405 \\
WTP20aanifd & 307.137 & -25.877 & 0.6049 & 0.0400 \\
WTP20aalcbh & 346.185 & -6.428 & 1.3229 & 0.0449 \\
WTP18aajsxw & 68.012 & -10.402 & 0.8242 & 0.0708 \\
WTP18aampwj & 26.677 & 32.508 & 4.6341 & 0.0375 \\
WTP20aajhyv & 219.582 & -26.292 & 1.1623 & 0.0473 \\
WTP14acrubs & 23.274 & -52.001 & 1.3283 & 0.0200 \\
WTP14acskaw & 20.976 & -35.065 & 24.2546 & 0.0191 \\
WTP14adbkno & 167.493 & 37.136 & 1.5988 & 0.0260 \\
WTP14adbjsh & 342.953 & -20.608 & 51.2272 & 0.0106 \\
WTP14aczser & 120.428 & 42.005 & 1.3214 & 0.0317 \\
WTP14acyjnm & 87.365 & -15.118 & 2.3063 & 0.0403 \\
WTP14acwadj & 175.035 & -76.554 & 1.2462 & 0.0403 \\
WTP18aamkzk & 237.180 & 22.137 & 6.9732 & 0.0313 \\
WTP15abymdq & 57.220 & -55.427 & 2.6354 & 0.0374 \\
WTP14acnjbu & 264.329 & 19.292 & 1.2734 & 0.0391 \\
WTP14adahfj & 317.000 & -29.836 & 2.6821 & 0.0199 \\
WTP14acznqs & 117.966 & 49.814 & 4.2951 & 0.0244 \\
WTP14ackmmo & 190.543 & -22.472 & 3.1717 & 0.0462 \\
WTP14acqpmt & 338.126 & -25.662 & 4.2604 & 0.0192 \\
WTP14acksqa & 64.394 & -52.885 & 0.7469 & 0.0456 \\
WTP14ackjvh & 191.489 & -12.856 & 0.9177 & 0.0781 \\
WTP14acmcha & 242.755 & -1.742 & 2.0505 & 0.0426 \\
WTP15abxvyr & 347.334 & 0.757 & 1.7067 & 0.0324 \\
WTP14adatfn & 173.483 & 67.019 & 1.5970 & 0.0397 \\
WTP17aalzpx & 50.857 & -2.047 & 1.5484 & 0.0372 \\
WTP14ackwxx & 217.085 & -0.204 & 3.8298 & 0.0241 \\
WTP19aalqow & 233.165 & 44.534 & 0.7117 & 0.0376 \\
WTP16aaveqn & 17.747 & 0.434 & 0.4767 & 0.0188 \\
WTP14acyxyr & 106.645 & 63.849 & 6.9545 & 0.0142 \\
WTP14acnfwn & 261.268 & 11.871 & 6.5440 & 0.0180 \\
WTP15aayqdd & 315.732 & 63.607 & 5.5862 & 0.0110 \\
WTP14acknzv & 214.498 & 25.137 & 14.6646 & 0.0172 \\
WTP19aaleui & 342.377 & 17.434 & 0.5336 & 0.0453 \\
WTP14adaqlx & 139.377 & 27.344 & 0.8947 & 0.0993 \\
WTP16aavohn & 298.911 & 61.512 & 0.6882 & 0.0641 \\
WTP18aanjtn & 332.060 & -73.167 & 1.9558 & 0.0465 \\
WTP18aajkmk & 40.099 & -2.728 & 4.9069 & 0.0289 \\
WTP17aaldjb & 197.065 & 4.486 & 0.5330 & 0.0483 \\
WTP18aaojju & 162.141 & -39.044 & 1.9341 & 0.0447 \\
WTP20aakleq & 165.412 & 28.844 & 0.3388 & 0.0344 \\
WTP15acbgpn & 166.258 & 59.685 & 4.1321 & 0.0337 \\
WTP18aajqqd & 232.965 & 37.413 & 0.3922 & 0.0299 \\
WTP16aavawr & 169.933 & 7.592 & 2.3399 & 0.0393 \\
WTP17aamzew & 146.236 & 31.098 & 1.1912 & 0.0347 \\
WTP14acrnct & 181.124 & 20.316 & 3.9250 & 0.0226 \\
WTP14acsdqt & 351.841 & 15.410 & 2.4538 & 0.0460 \\
WTP14acvhdx & 36.427 & 31.623 & 0.9117 & 0.0172 \\
WTP15acbfkt & 162.040 & 56.917 & 0.3062 & 0.0462 \\
WTP14adeqka & 297.354 & 63.509 & 12.7971 & 0.0190 \\
WTP14ackodc & 207.025 & 7.388 & 1.2301 & 0.0235 \\
WTP13aabqed & 350.530 & 10.306 & 0.9033 & 0.0612 \\
WTP15abyuwj & 57.598 & -50.310 & 0.8098 & 0.0365 \\
WTP14acttuy & 202.808 & -25.403 & 7.6342 & 0.0260 \\
WTP18aakapq & 247.998 & 24.628 & 0.2943 & 0.0433 \\
WTP14acpcnd & 128.912 & -4.088 & 7.7089 & 0.0142 \\
WTP19aakidi & 285.670 & 48.619 & 0.4700 & 0.0608 \\
WTP17aanbso & 180.910 & 58.987 & 0.1728 & 0.0469
\label{table:flares}
\end{longtable}

\end{document}